\documentclass[english,american,round]{article}
\usepackage[T1]{fontenc}
\usepackage[a4paper]{geometry}
\usepackage{fancyhdr}
\usepackage{babel}
\usepackage{prettyref}
\usepackage{mathtools}
\usepackage{amsmath}
\usepackage{amsthm}
\usepackage{amssymb}
\usepackage{graphicx}
\usepackage{setspace}
\usepackage{textcomp,marvosym}
\usepackage{wasysym}
\usepackage{esint}
\usepackage[authoryear]{natbib}
\usepackage{xargs}[2008/03/08]
\usepackage[unicode=true,
 bookmarks=false,bookmarksnumbered=false,bookmarksopen=false,
 breaklinks=true,pdfborder={0 0 0},backref=false]
 {hyperref}
\hypersetup{pdftitle={Fundamental activity constraints lead to specific interpretations of the connectome}, 
colorlinks=true,linkcolor=black,citecolor=black,urlcolor=black,filecolor=black,
 pdfauthor={Jannis Schuecker, Maximilian Schmidt, Sacha van Albada, Markus Diesmann, Moritz Helias}}
\usepackage{breakurl}

\makeatletter
\usepackage{tabularx} 
\usepackage{multirow}
\usepackage{colortbl}
\usepackage{dsfont}

\usepackage{prettyref}
\usepackage{ifthen}
\newcommand{\arXiv}{true}
\newboolean{isarxiv}
\setboolean{isarxiv}{true}
\newrefformat{cap}{\hyperref[#1]{Fig.~\ref{#1}}}
\newrefformat{fig}{\hyperref[#1]{Fig.~\ref{#1}}}
\newrefformat{tab}{\hyperref[#1]{Table ~\ref{#1}}}
\newrefformat{sec}{\hyperref[#1]{Sec.~\ref{#1}}}
\newrefformat{sub}{\hyperref[#1]{Sec.~\ref{#1}}}

\usepackage{braket}

\usepackage[OT1]{fontenc}

\usepackage{tocbibind}

\settocbibname{References}

\renewcommand{\@seccntformat}[1]{%
\csname the#1\endcsname\hspace{0.5em}}

\let\size@subsection\LARGE

\ifthenelse{\equal{\arXiv}{true}}{

\renewcommand{\@makecaption}[2]{%
{\parbox[t]{\linewidth}{%
\normalsize\renewcommand{\baselinestretch}{1.0}\normalsize
\vspace{2mm}
\textsf{\textbf{#1} #2}
}}}

}
{ 
\usepackage[tabhead,nofighead,tablesfirst]{endfloat}

\AtBeginDelayedFloats{%
\sffamily%
\linespread{2}\selectfont%
}

\renewcommand{\@makecaption}[2]{%
{\parbox[t]{\linewidth}{%
\normalsize\renewcommand{\baselinestretch}{1.0}\normalsize
\vspace{2mm}
\centering
\textbf{\textsf{#1}}
}}}
}

\usepackage{lineno}

\@ifundefined{definecolor}
 {\usepackage{color}}{}

\definecolor{parametergray}{gray}{0.8}

\makeatother

\begin{document}
\global\long\def\A{\mathbf{A}}

\global\long\def\A{{\text{A}^{-}}}

\global\long\def\a{\mathbf{a}}

\global\long\def\B{\mathbf{B}}

\global\long\def\b{\mathbf{b}}

\global\long\def\C{\mathbf{C}}

\global\long\def\c{\mathbf{c}}

\global\long\def\Ca{{\text{Ca}^{2+}}}

\global\long\def\cc{{^{*}}}

\global\long\def\Cl{{\text{Cl}^{-}}}

\global\long\def\Cov#1{\text{Cov}\left(#1\right)}

\global\long\def\d{\mathrm{d}}

\global\long\def\diff#1#2{{\displaystyle \frac{\text{d}#1}{\text{d}#2}}}

\global\long\def\D{\mathbf{D}}

\newcommandx\EW[2][usedefault, addprefix=\global, 1=]{\left\langle #2\right\rangle _{#1}}

\selectlanguage{english}%
\global\long\def\EL{E_{\mathrm{L}}}

\selectlanguage{american}%
\global\long\def\e{\mathbf{e}}

\global\long\def\erf{\mathrm{erf}}

\global\long\def\erfc{\mathrm{erfc}}

\global\long\def\Em{\mathbf{1}}

\global\long\def\Ex{\mathcal{E}}

\global\long\def\diag{\mathrm{diag}}

\global\long\def\E{\text{E}}

\global\long\def\ftr{\mathcal{F}}

\global\long\def\f{\mathbf{f}}

\global\long\def\Ftr#1#2{\mathfrak{F}[#1](#2)}

\global\long\def\iFtr#1#2{\mathfrak{F}^{-1}[#1](#2)}

\global\long\def\h{\mathbf{h}}

\global\long\def\G{\mathbf{G}}

\global\long\def\Hz{\:\mathrm{Hz}}

\global\long\def\In{\mathcal{I}}

\global\long\def\inp{\text{inp}}

\global\long\def\I{\text{I}}

\global\long\def\Int#1#2#3#4{\int\limits _{#3}^{#4} \text{d}#2\ #1}

\global\long\def\j{\mathbf{j}}

\global\long\def\J{\mathbf{J}}

\global\long\def\K{{\text{K}^{+}}}

\global\long\def\LN{\mathcal{L}_{0}}

\global\long\def\LO{\mathcal{L}_{1}}

\global\long\def\Lfp{L_{\mathrm{FP}}}

\global\long\def\M{\mathbf{M}}

\global\long\def\m{\mathbf{m}}

\global\long\def\mus{\:\mu\mathrm{s}}

\global\long\def\ms{\,\text{ms}}

\global\long\def\mV{\,\text{mV}}

\selectlanguage{english}%
\global\long\def\pA{\:\mathrm{pA}}

\global\long\def\pF{\:\mathrm{pF}}

\global\long\def\cm{C_{\mathrm{m}}}

\selectlanguage{american}%
\global\long\def\N{\mathbf{N}}

\global\long\def\n{\mathbf{n}}

\global\long\def\Na{{\text{Na}^{+}}}

\global\long\def\nuo{\nu_{0}}

\global\long\def\nr{n_{r}}

\global\long\def\OO{\mathbf{O}}

\global\long\def\P{\mathbf{P}}

\global\long\def\PV{P(V)}

\global\long\def\pF{\:\mathrm{pF}}

\global\long\def\Q{\mathbf{Q}}

\global\long\def\q{\mathbf{q}}

\global\long\def\R{\mathbf{R}}

\global\long\def\r{\mathbf{r}}

\global\long\def\res{\mathrm{Res}}

\global\long\def\rest{\text{rest}}

\global\long\def\s{\mathbf{s}}

\global\long\def\sps{\text{s}^{-1}}

\global\long\def\tilPV{\tilde{P}(V)}

\global\long\def\v{\mathbf{v}}

\global\long\def\V{\mathbf{V}}

\global\long\def\Var{\mathrm{Var}}

\global\long\def\Vth{V_{\theta}}

\global\long\def\Vr{V_{r}}

\global\long\def\x{\mathbf{x}}

\global\long\def\y{\mathbf{y}}

\global\long\def\T{\mathbf{T}}

\global\long\def\transp{\mathbf{^{\text{T}}}}

\global\long\def\uvec{\boldsymbol{u}}

\global\long\def\U{\mathbf{U}}

\global\long\def\Var#1{\text{Var}\left(#1\right)}

\global\long\def\vec#1{\boldsymbol{#1}}

\global\long\def\w{\boldsymbol{w}}

\global\long\def\W{\mathbf{W}}

\global\long\def\X{\mathbf{X}}

\global\long\def\y{\mathbf{y}}

\global\long\def\Y{\mathbf{Y}}

\global\long\def\taue{\tau_{e}}

\global\long\def\taum{\tau_{\mathrm{m}}}

\global\long\def\taus{\tau_{\mathrm{s}}}

\global\long\def\taur{\tau_{\mathrm{r}}}

\global\long\def\tauM{\tau_{\text{m}}}

\global\long\def\tauR{\tau_{\text{ref}}}

\global\long\def\nuext{\nu_{\mathrm{ext}}}

\global\long\def\defeq{\vcentcolon=}

\global\long\def\rdefeq{\mathrm{=\vcentcolon}}

\global\long\def\orderofmagnitude{\sim}

\global\long\def\unity{\mathds{1}}

\global\long\def\spikess{\;\mathrm{spikes}/\mathrm{s}}

\global\long\def\tav#1{\widehat{#1}}

\global\long\def\av#1{\overline{#1}}

\global\long\def\dotv{\mathord\cdot}

\newcommand{\runningtitle}{Activity constraints on the connectome}

\begin{titlepage}\thispagestyle{empty}\setcounter{page}{0}\pdfbookmark[1]{Title}{TitlePage}

\noindent \begin{center}
\textbf{\huge{}Fundamental activity constraints lead to specific interpretations
of the connectome}
\par\end{center}{\huge \par}

\noindent \begin{center}
{\large{}\textbf{Jannis Schuecker}\textsuperscript{1,\Yinyang}, \textbf{Maximilian Schmidt} \textsuperscript{1,\Yinyang},
\textbf{Sacha J. van Albada}\textsuperscript{1}, \textbf{Markus Diesmann}\textsuperscript{1,2,3}, \textbf{and Moritz
Helias}\textsuperscript{1,3}}
\par\end{center}{\large \par}

\vspace{2cm}

\noindent$^{1}$\parbox[t]{15cm}{Institute of Neuroscience and Medicine
(INM-6) and Institute for Advanced Simulation (IAS-6) and JARA BRAIN
Institute I, J\"ulich Research Centre, 52428 J\"ulich, Germany}\\[3mm]

\noindent$^{2}$\parbox[t]{15cm}{Department of Psychiatry, Psychotherapy
and Psychosomatics, Medical Faculty, RWTH Aachen University, 52062
Aachen, Germany}\\[3mm]

\noindent$^{3}$\parbox[t]{15cm}{Department of Physics, Faculty
1, RWTH Aachen University, 52062 Aachen, Germany}\\[3mm]

\noindent \textsuperscript{\Yinyang}\parbox[t]{15cm}{Both authors contributed equally
to this work.}\\[3mm]\noindent

\vfill

\ifthenelse{\equal{\arXiv}{true}}{}{

\noindent \parbox[t]{15cm}{\textbf{Running title:} \runningtitle}\\

\noindent \parbox[t]{15cm}{\textbf{Statistics:}\\
Number of pages: 34\\
Number of figures: 9\\
Number of tables: 0\\
Number of words in abstract: 201\\
Number of words in introduction: 642\\
Number of words in discussion: 1500}

\vspace{2cm}
}

\noindent Correspondence to:\hspace{1em}\parbox[t]{11cm}{Jannis
Schuecker, Wilhelm-Johnen-Stra\upshape{\ss}e, 52425 J\"ulich\\
\href{mailto:j.schuecker@fz-juelich.de}{j.schuecker@fz-juelich.de}}

\lhead{\runningtitle}

\rhead{Schuecker, Schmidt et al.}

\end{titlepage}

\section{Abstract }

The continuous integration of experimental data into coherent models
of the brain is an increasing challenge of modern neuroscience. Such
models provide a bridge between structure and activity, and identify
the mechanisms giving rise to experimental observations. Nevertheless,
structurally realistic network models of spiking neurons are necessarily
underconstrained even if experimental data on brain connectivity are
incorporated to the best of our knowledge. Guided by physiological
observations, any model must therefore explore the parameter ranges
within the uncertainty of the data. Based on simulation results alone,
however, the mechanisms underlying stable and physiologically realistic
activity often remain obscure. We here employ a mean-field reduction
of the dynamics, which allows us to include activity constraints into
the process of model construction. We shape the phase space of a multi-scale
network model of the vision-related areas of macaque cortex by systematically
refining its connectivity. Fundamental constraints on the activity,
i.e., prohibiting quiescence and requiring global stability, prove
sufficient to obtain realistic layer- and area-specific activity.
Only small adaptations of the structure are required, showing that
the network operates close to an instability. The procedure identifies
components of the network critical to its collective dynamics and
creates hypotheses for structural data and future experiments. The
method can be applied to networks involving any neuron model with
a known gain function.

\section{Author summary}

The connectome describes the wiring patterns of the neurons in the
brain, which form the substrate guiding activity through the network.
The influence of its constituents on the dynamics is a central topic
in systems neuroscience. We here investigate the critical role of
specific structural links between neuronal populations for the global
stability of cortex and elucidate the relation between anatomical
structure and experimentally observed activity. Our novel framework
enables the evaluation of the rapidly growing body of connectivity
data on the basis of fundamental constraints on brain activity and
the combination of anatomical and physiological data to form a consistent
picture of cortical networks.

\ifthenelse{\boolean{isarxiv}}{}{\linenumbers}

\section{Introduction}

The neural wiring diagram, the connectome, is gradually being filled
through classical techniques combined with innovative quantitative
analyses \citep{Markov2014_17,Markov14} and new technologies \citep{Wedeen08_1267,Axer11}.
The connectivity between neurons is considered to shape resting-state
and task-related collective activity \citep{Cole2014_238,Shen15_5579}.
For simple networks, clear relationships with activity are known analytically,
e.g., a dynamic balance between excitatory and inhibitory inputs in
inhibition-dominated random networks leads to an asynchronous and
irregular state \citep{Vreeswijk96,Amit-1997_373,Tetzlaff12_e1002596}.
Structures and mechanisms underlying large-scale interactions have
been identified by means of dynamical models \citep{Deco12_3366,cabral14_102}
and graph theory \citep{Markov13_Science,Goulas14_e1003529}. Furthermore,
the impact of local network structure on spike-time correlations is
known in some detail \citep{Ostojic09_10234,Pernice11_e1002059,Trousdale12_e1002408}.
Under certain conditions, there is a one-to-one mapping between correlations
in neuronal network activity and effective connectivity, a measure
that depends on the network structure and on its activity \citep{Albada15}.
In conclusion, anatomy does not uniquely determine dynamics, but dynamical
observations help constrain the underlying anatomy. Despite advances
in understanding simple networks, a complete picture of the relationship
between structure and dynamics remains elusive.

\begin{figure}
\begin{centering}
\includegraphics{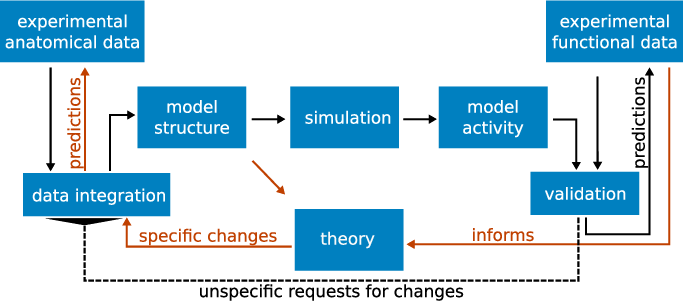}
\par\end{centering}
\caption{\textbf{The integrative loop between experiment, model, and theory.}
Black arrows represent the classical forward modeling approach: Experimental
anatomical data are integrated into a model structure, which gives
rise to the activity via simulation. The model activity is compared
with experimental functional data. The usual case of disagreement
leads to the need to change the model definition. By experience and
intuition the researcher identifies the parameters to be changed,
proceeding in an iterative fashion. Once the model agrees well with
a number of experimental findings, it may also generate predictions
on the outcome of new experiments. Red arrows symbolize our new approach:
informed by functional data, an analytical theory of the model identifies
critical connections to be modified, refining the integration of data
into a consistent model structure and generating predictions for anatomical
studies. }

\label{fig:scheme}
\end{figure}

\prettyref{fig:scheme} visualizes the integrative loop between experiment,
model, and theory that guides the investigation of structure and dynamics.
In the traditional forward modeling approach, structural data from
experimental studies determine the connectivity between model neurons.
Combined with the specification of the single-neuron dynamics and
synaptic interactions, simulations yield a certain network dynamics.
There is a fundamental problem with this approach.

Despite the impressive experimental progress in determining network
parameters, any neural network model is underdetermined, because of
the sheer complexity of brain tissue and inevitable uncertainties
in the data. For instance, counting synapses on a large scale presently
takes a prohibitive amount of time, and no available technique allows
determining precise synaptic weights for entire neural populations.
Although it may be possible to quantify the full connectome of an
individual in future, inter-individual variability will require modelers
to express connectivity as connection probabilities or rules of self-organization
if they want to learn about general principles of the brain. In modeling
studies, parameters are usually tuned manually to achieve a satisfactory
state of activity, which becomes unfeasible for high-dimensional models
due to the size of the parameter space. In particular, it is a priori
unclear how parameters of the model influence its activity. In consequence,
modifications cannot be performed in a targeted fashion, and it is
difficult to find a minimal set of modifications necessary for aligning
a model with experimental activity data.

Overcoming this problem requires a shift of perspective. Instead of
regarding the model exclusively in a forward manner, generating predicted
activity from structure, we in addition consider the system in a reverse
manner, predicting the structure necessary to explain the observed
activity. Our theory, starting from a mean-field description, provides
a direct link between structure and activity. In contrast to simulations,
the theory is invertible, which we exploit to identify connections
critical for the dynamics and to find a minimal set of modifications
to the structure yielding a realistic set point of activity. The predicted
alterations generate hypotheses on brain structure, thus feeding back
to anatomical experiments.

Applying the method to a multi-scale network model of the vision-related
areas of macaque cortex, we derive targeted modifications of a set
of critical connections, bringing the model closer to experimental
observations of cortical activity. Based on the global properties
of the bistable phase space of the model, the method refines the model's
construction principles within experimental uncertainties and identifies
the connections that decisively shape the dynamics. Preliminary results
have been presented in abstract form \citep{Schmidt15_T26-7C}.

\section{Results\label{sec:Results}}

\subsection{Global stability in a simple network}

We consider networks with neurons structured into $N$ interconnected
populations. A neuron in population $i$ receives $K_{ij}$ incoming
connections from neurons in population $j$, each with synaptic efficacy
$J_{ij}$. Additionally, each neuron is driven by $K_{\mathrm{ext}}$
Poisson sources with rate $\nuext$ and synaptic efficacy $J_{\mathrm{ext}}$.
All neurons in one population have identical parameters, so that we
can describe the network activity in terms of population-averaged
firing rates $\nu_{i}$.

Our method is applicable if the employed neuron model has a known
gain function, either analytically or as a function obtained by interpolating
numerical results from simulations. In this study, we model single
cells as leaky integrate-and-fire model (LIF) neurons (see \nameref{sec:Theory-and-Methods}).
The possible stationary states of these networks are characterized
by the firing rates that are equilibria of

\begin{equation}
\dot{\boldsymbol{\nu}}\coloneqq\frac{\d\boldsymbol{\nu}}{\d s}=\boldsymbol{\Phi}(\boldsymbol{\nu},\boldsymbol{A})-\boldsymbol{\nu},\label{eq:int_siegert}
\end{equation}
where $s$ is a pseudo-time. The gain function $\boldsymbol{\Phi}$
is known analytically and $\boldsymbol{A}$ indicates its dependence
on the model parameters $\{\boldsymbol{K},\boldsymbol{J},\nuext,\dots\}$
(see \nameref{sec:Theory-and-Methods}).

\begin{figure}
\begin{centering}
\includegraphics{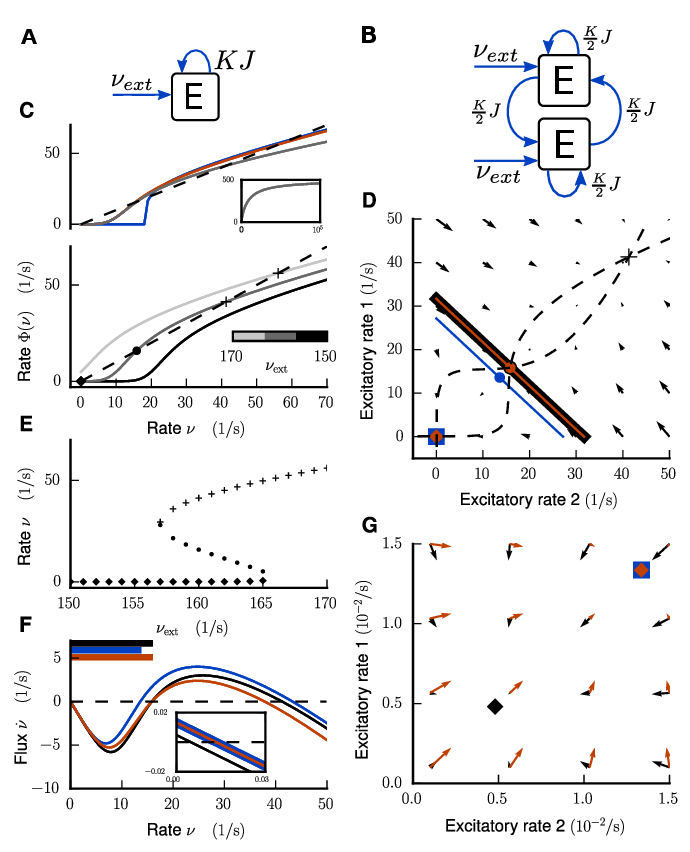}
\par\end{centering}
\caption{\textbf{Activity flow in an illustrative network example.} \textbf{Left
column:} Global stability analysis in the single-population network.
\textbf{A} Illustration of network architecture. \textbf{C} Upper
panel: Input-output relationship $\Phi(\nu,\protect\nuext)$ for external
Poisson drive $\protect\nuext=160\,\frac{1}{\mathrm{s}}$ shown in
gray. In addition $\Phi(\nu,\protect\nuext)$ for $\protect\taur=0$
for the noiseless case (blue) and the noisy case (red). The inset
shows the gray curve over a larger input range. Lower panel: $\Phi(\nu,\protect\nuext)$
for different rates of the external Poisson drive $\protect\nuext=[150,160,170]\,\frac{1}{\mathrm{s}}$
from black to light gray. Intersections with the identity line (dashed)
mark fixed points of the system, which are shown in \textbf{E} as
a function of $\protect\nuext$. \textbf{F} Flux $\dot{\nu}$ in the
bistable case for $\Phi(\protect\nuext=160\frac{1}{\mathrm{s}},K)$
in black, $\Phi(\protect\nuext^{\prime}=161\frac{1}{\mathrm{s}},K)$
in blue, and modified system $\Phi(\protect\nuext^{\prime}=161\frac{1}{\mathrm{s}},K^{\prime})$
in red. Intersections with zero (dashed line) mark fixed points. The
inset shows an enlargement close to the LA fixed point. Horizontal
bars at top of figure denote the size of the basin of attraction for
each of the three settings. \textbf{Right column:} Global stability
analysis in the network of two mutually coupled excitatory populations.\textbf{
B }Illustration of network architecture.\textbf{ D} Flow field and
nullclines (dashed curves) for $\boldsymbol{\Phi}(\protect\nuext=160\frac{1}{\mathrm{s}},K)$
and separatrices (solid lines), LA fixed point (rectangle), HA fixed
point (cross) and unstable fixed points (circles) for $\boldsymbol{\Phi}(\protect\nuext=160\frac{1}{s},K)$
in black, $\boldsymbol{\Phi}(\protect\nuext=161\frac{1}{s},K)$ in
blue, and $\boldsymbol{\Phi}(\protect\nuext=161\frac{1}{s},K^{\prime})$
in red. The red separatrix and the red unstable fixed point coincide
with the black ones. \textbf{G} Enlargement of D close to the LA fixed
points. Flow field of original system shown in black, of modified
system in red.}
\label{fig:Brunel-2D}
\end{figure}

The input-output relationship $\boldsymbol{\Phi}$ typically features
a non-linearity which, in combination with feedback connections, can
cause multi-stability in the network. In particular excitatory-excitatory
loops cause the system defined by \prettyref{eq:int_siegert} to exhibit
multi-stable behavior in the stationary firing rates. A necessary
condition for the bistability is that the transfer function has an
inflection point. The LIF neuron model can have such an inflection
point, originating from the interplay of its leak term and the threshold.
Less realistic neuron models, such as the perfect integrate-and-fire
model, do not have such an inflection point. To illustrate its origin,
we first consider the noiseless case \citep{Dayan01} without absolute
refractoriness ($\taur=0$). The transfer function initially grows
from zero with infinite slope due to the threshold and crosses the
identity line (\prettyref{fig:Brunel-2D}C). For larger input the
leak term can be neglected and the transfer function approaches a
linear function with finite slope $\frac{1}{\taum}\,\frac{R}{V_{\theta}-V_{\mathrm{r}}}$ 
 (see, e.g., \citep{Kriener14}, eq. 11), equivalent to a perfect
integrator. This is only possible with a negative curvature at intermediate
rates, i.e., a reduction in the slope, which makes the transfer function
cross the identity line another time, causing the bistability. Network-generated
noise only affects the low-rate regime where it smears out the kink
causing the transfer function to grow from zero with positive curvature
(see, e.g., \citep{Brunel00}, eq. 22). Importantly, the qualitative
picture, i.e., the bistable behavior, is not altered. A finite refractory
period only has an effect for very high input values where the transfer
function saturates at $1/\taur=500\spikess$ for the given parameters.

The basic problem is that there is a trade-off between excitation
at the fixed points and their stability. In particular, exciting the
model to bring a fixed point closer to experimental observations requires
a method to preserve its stability. We achieve this by controlling
the influence of the excitatory-excitatory loops on the phase space
of the network.

As an illustration we first study the mechanism using the simple network
architecture depicted in \prettyref{fig:Brunel-2D}A: a population
of excitatory neurons is coupled to itself with indegree $K$ and
is driven by external Poisson sources with the same indegree $K_{\mathrm{ext}}=K$
and rate $\nu_{\mathrm{ext}}$. All connections have identical synaptic
weights $J$. An increase in the external drive shifts the input-output
relationship $\Phi(\nu,\nu_{\mathrm{ext}})$ of this one-dimensional
system (\prettyref{fig:Brunel-2D}C) to the left. The bifurcation
diagram is shown in \prettyref{fig:Brunel-2D}E: for low $\nu_{\mathrm{ext}}$
there is only one fixed point with low activity (LA). When increasing
$\nu_{\mathrm{ext}}$, an additional pair of fixed points of which
one is stable and the other is unstable emerges via a saddle-node
bifurcation, leading to a bistable system. The second stable fixed
point exhibits high firing rates, denoted as the high-activity (HA)
state. For even higher values of $\nu_{\mathrm{ext}}$, the LA state
loses stability.

The equilibria, given by the zeros of the velocity $\dot{\nu}$ in
the bistable case, are shown in \prettyref{fig:Brunel-2D}F. An increase
of the drive on the one hand increases the firing rate of the LA fixed
point (inset) but on the other hand reduces its basin of attraction,
indicated by the colored bars in the top left corner. For illustrative
purposes, we extend the problem to two dimensions by splitting the
excitatory population into two subpopulations of equal size (\prettyref{fig:Brunel-2D}B),
mimicking a loop between excitatory populations in the model of the
vision-related areas of macaque cortex. The corresponding (symmetric)
two-dimensional phase space is shown in \prettyref{fig:Brunel-2D}D.
The basin of attraction for the LA fixed point, limited by the separatrix
\citep{Strogatz94}, is reduced with increasing external drive. 

Since we have a bistable system, there must be at least one unstable
fixed point on the separatrix at the intersection of the nullclines,
i.e., the subspace for which the velocity $\dot{\nu}_{i}$ in direction
$i$ vanishes. We use the unstable fixed point to preserve the basin
of attraction when the external drive $\nuext$ is increased. For
this purpose, we modify the connectivity $K\rightarrow K^{\prime}$
to reverse the shift of the unstable fixed point due to the parameter
change $\nuext\rightarrow\nuext^{\prime}$ (see \nameref{sec:Theory-and-Methods}
for a detailed derivation). Since the separatrix follows the unstable
fixed point, this approximately restores the original basin of attraction.

The resulting velocity of the system $\Phi(\nuext^{\prime},K^{\prime})$
($\Phi$ defines the system \prettyref{eq:int_siegert}) is shown
in \prettyref{fig:Brunel-2D}F. The firing rate in the LA fixed point
is increased as desired (inset), and the unstable fixed point coincides
with that obtained in the original system $\Phi(\nuext,K)$. This
pattern of fixed points is also indicated by the zero vectors of the
velocity field $\boldsymbol{\dot{\nu}}$ (\prettyref{fig:Brunel-2D}D).
The separatrix follows the unstable fixed point, and the basin of
attraction in the system $\boldsymbol{\Phi}(\nuext^{\prime},K^{\prime}$)
is restored to that in the original system $\boldsymbol{\Phi}(\nuext,K$).
\prettyref{fig:Brunel-2D}G shows the behavior of the LA fixed point
in more detail. The modification of $K$ does not noticeably change
the location of the LA fixed point. In conclusion, the method allows
us to increase the firing rates in the LA fixed point without modifying
its basin of attraction.

The purely excitatory network is the simplest model to explain a phase
space configuration with a LA and a HA fixed point. Inhibitory feedback
is not necessary for this bistability, but it would certainly alter
the input-output relationship. For example the classical excitatory-inhibitory
network \citep{Brunel00} in the balanced regime has an input-output
relationship with a negative slope and thus only one fixed point exists.
However, if a pair of such balanced E-I networks is coupled by sufficiently
strong mutually excitatory connections, these connections cause a
bistability in a similar manner. Thus the mechanism shown in the purely
excitatory network can also lead to the emergence of a HA attractor
in more complex networks with inhibition.

\subsection{Bistability in the multi-scale network model}

\begin{figure}
\begin{centering}
\includegraphics{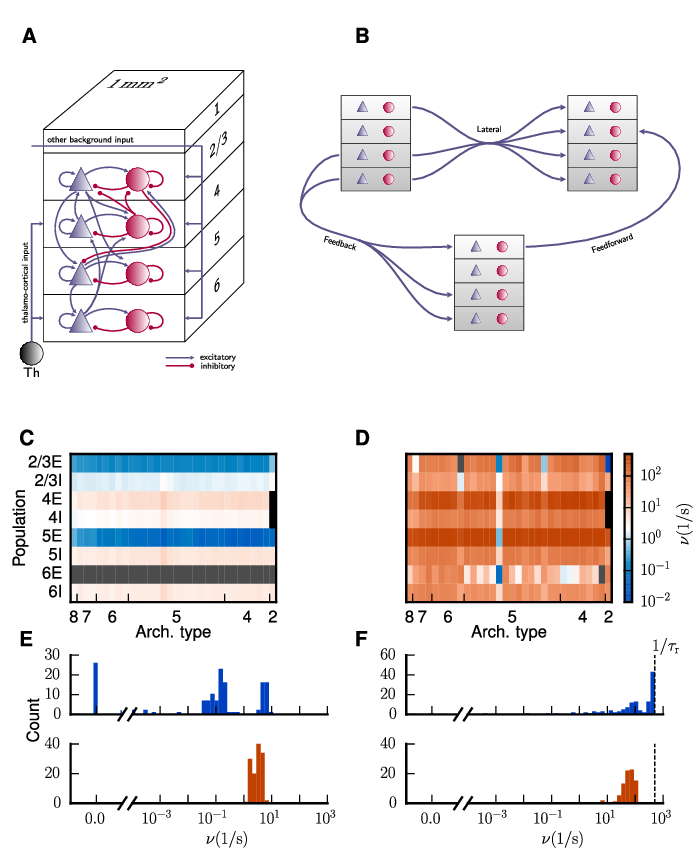}
\par\end{centering}
\centering{}\caption{\textbf{Bistability of the model. A} Sketch of the microcircuit model
serving as a prototype for the areas of the multi-area model (figure
and legend adapted from figure 1 of Potjans and Diesmann \citep{Potjans14_785},
with permission).\textbf{ B }Sketch of the most common laminar patterns
of cortico-cortical connectivity of the multi-area model. \textbf{C}
Population-averaged firing rates encoded in color for a spiking network
simulation of the multi-area model with low external drive ($\kappa=1.0$).
\textbf{D }As\textbf{ }C but for increased external drive ($\kappa=1.125$).
The color bar refers to both panels. Areas are ordered according to
their architectural type along the horizontal axis from V1 (type 8)
to TH (type 2) and populations are stacked along the vertical axis.
The two missing populations 4E and 4I of area TH are marked in black
and firing rates $<10^{-2}\protect\Hz$ in gray. \textbf{E} Histogram
of population-averaged firing rates shown in \textbf{C} for excitatory
(blue) and inhibitory (red) populations. The horizontal axis is split
into linear- (left) and log-scaled (right) ranges. \textbf{F} as E
corresponding to state shown in D. }
\label{fig:bistability-multiarea}
\end{figure}

We investigate a multi-scale network model of cortical areas to understand
the structural features essential for a realistic state of baseline
activity. The model extends and adapts the microcircuit model presented
in \citep{Potjans14_785}, which covers $1\:\mathrm{mm}^{2}$ surface
area of early sensory cortex (\prettyref{fig:bistability-multiarea}A),
to all vision-related areas of macaque cortex (\prettyref{fig:bistability-multiarea}B).
Based on the microcircuit model, an area is composed of 4 layers (2/3,
4, 5, and 6) each having an excitatory (E) and an inhibitory (I) neural
population, except parahippocampal area TH, which consists of only
3 layers (2/3, 5, and 6). A detailed description of the data integration
is given in \citep{Schmidt16_arxiv_v4}.

Simulations of the model (\prettyref{fig:bistability-multiarea}C)
reveal that, though realistic levels of activity can be achieved for
populations in layers 2/3 and 4, populations 5E and 6E of the majority
of areas show vanishingly low or zero activity in contrast to empirical
data \citep{Swadlow88_1162,deKock09_16446}. Inputs from subcortical
and non-visual cortical areas are modeled as Poissonian spike trains,
whose rate $\nuext$ is a free, global parameter. To elevate the firing
rates in the excitatory populations in layers 5 and 6, we enhance
the external Poisson drive onto these populations parametrized by
$\kappa$ (see \nameref{sec:Theory-and-Methods}). However, already
a perturbation of a few percent leads to a state with unrealistically
high rates (\prettyref{fig:bistability-multiarea}D), caused by the
reduced basin of attraction of the low-activity state similar to \prettyref{fig:Brunel-2D}D.
Our aim is to improve the working point of the model such that all
populations exhibit spiking activity $\apprge0.05\spikess$ while
preventing the model from entering a state with unrealistically high
rates of $\apprge30\spikess$ (figure 13 of \citep{Swadlow88_1162},
\citep{deKock09_16446}). The employed technique exposes the mechanism
giving rise to the observed instability and identifies the circuitry
responsible for this dynamical feature.

\subsection{Targeted modifications preserve global stability}

We apply the procedure derived in \nameref{sec:Theory-and-Methods}
and find targeted modifications to the connectivity $\boldsymbol{K}$
that preserve the global stability of the low-activity fixed point
for increased values of the external drive $\kappa$.

\begin{figure}
\begin{centering}
\includegraphics{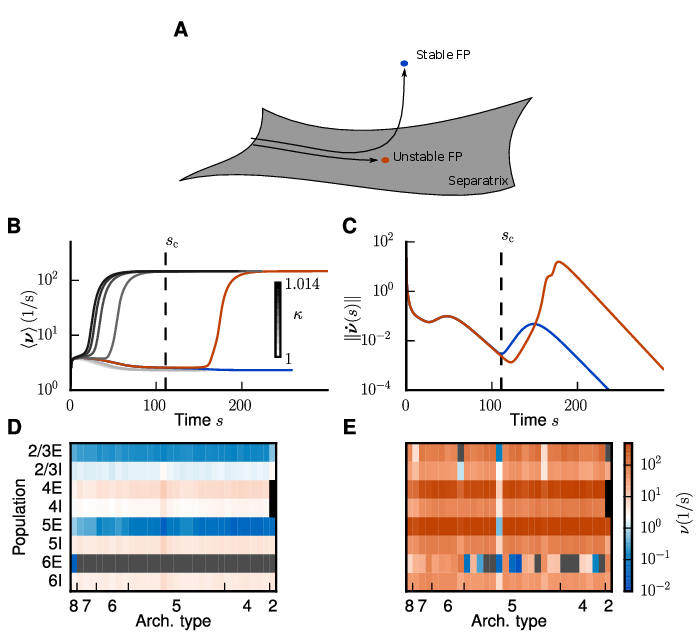}
\par\end{centering}
\centering{}\caption{\textbf{Application of the mean-field theory to the multi-area model.
A} Trajectories of \prettyref{eq:int_siegert} starting inside the
separatrix converge to an unstable fixed point. Trajectories starting
close to the separatrix are initially attracted by the unstable fixed
point but then repelled in the fixed point's unstable direction and
finally converge to a stable fixed point.\textbf{ B} Firing rate averaged
across populations over time. Integration of \prettyref{eq:int_siegert}
leads to convergence to either the low-activity (LA) or the high-activity
(HA) attractor for different choices of the external input factor
$\kappa$, with $\kappa=1$ the original level of external drive.
We show eight curves with $\kappa$ varying from $1.0$ to $1.014$
in steps of $0.002$ and two additional curves for $\kappa=1.007662217,\,1.007662218$.
The curves for the largest factor ($\kappa=1.007662217$) that still
leads to the LA state and for the smallest factor ($\kappa=1.007662218$)
that leads to the HA state are marked in blue and red, respectively.
The four curves with $\kappa\leq1.006$ coincide with the blue curve.
\textbf{C} Euclidean norm of the velocity vector in the integration
of \prettyref{eq:int_siegert} for the different choices of $\kappa$.
The vertical dashed line indicates the time $s_{\mathrm{c}}$ of the
last local minimum in the blue curve. \textbf{D} Stationary firing
rate in the different areas and layers of the model in a low-activity
state for $\kappa=1.0$ as predicted by the mean-field theory (same
display as in \prettyref{fig:bistability-multiarea}). \textbf{E}
As D, but showing the high-activity state for $\kappa=1.125$.}
\label{fig:mf-multiarea}
\end{figure}

In the following we choose the inactive state $\boldsymbol{\nu}(0)=(0,\dots,0)^{\mathrm{T}}$
as the initial condition. The exact choice is not essential since
we are only interested in the fixed points of the system. \prettyref{fig:mf-multiarea}B
shows the integration of \prettyref{eq:int_siegert} over pseudo-time
$s$ for different levels of the external drive to populations 5E
and 6E parametrized by $\kappa$. For low values of $\kappa$, the
integration converges to the LA fixed point shown in \prettyref{fig:mf-multiarea}D,
and is in agreement with the activity emerging in the simulation (\prettyref{fig:bistability-multiarea}C).
For increased values of $\kappa$, the system settles in the HA fixed
point (\prettyref{fig:mf-multiarea}E), again in agreement with the
simulation. The population-specific firing rates in the HA state found
in the mean-field predictions (\prettyref{fig:mf-multiarea}E) are
also close to those in the simulation (\prettyref{fig:bistability-multiarea}D),
but minor deviations occur due to the violation of the assumptions
made in the diffusion approximation. In particular, at these pathologically
high rates, the neurons fire regularly and close to the reciprocal
of their refractory period, while in the mean-field theory we assume
Poisson spike statistics. Still, the mean-field theory predicts the
bistability found in the simulation. Since the theory yields reliable
predictions in both stable fixed points, we assume that also the location
of the unstable fixed point in between these two extremes is accurately
predicted by the theory.

\begin{figure}
\begin{centering}
\includegraphics{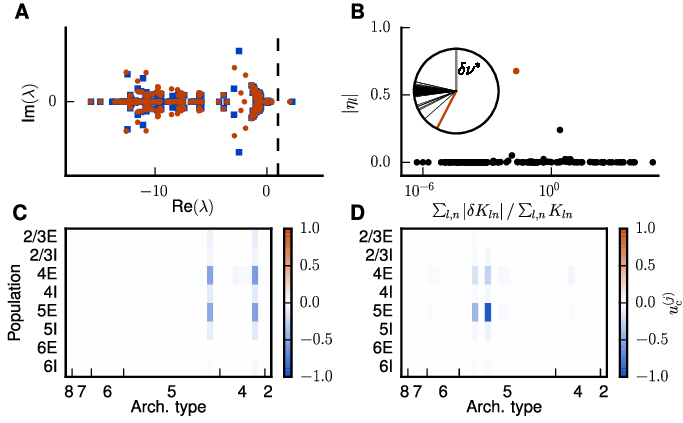}
\par\end{centering}
\centering{}\caption{\textbf{Eigenspectrum analysis of network stability. A} Eigenvalue
spectrum of the effective connectivity matrix $\boldsymbol{M}$ for
the first (blue squares) and second (red dots) iteration. The dashed
vertical line marks the edge of stability at a real part of 1. \textbf{B}
Contribution $\eta_{l}$ (\prettyref{eq:individual_shifts}) of an
individual eigenprojection $l$ to the shift of the unstable fixed
point versus the relative change in indegrees associated with $l$
for the first iteration. The data point corresponding to $\lambda_{\mathrm{c}}^{(1)}$
is marked in red. The inset shows the relative angles between $\boldsymbol{\delta\nu^{\ast}}$
and the eigenvectors $\boldsymbol{u}^{l}$. The red line corresponds
to the critical eigendirection. \textbf{C} Entries of the eigenvector
$u_{c}^{(1)}$ associated with $\lambda_{\mathrm{c}}^{(1)}$ in the
populations of the model. The affected areas are 46 and FEF. \textbf{D}
Same as C for the second iteration. }
\label{fig:eigenspace}
\end{figure}

\begin{figure}
\begin{centering}
\includegraphics{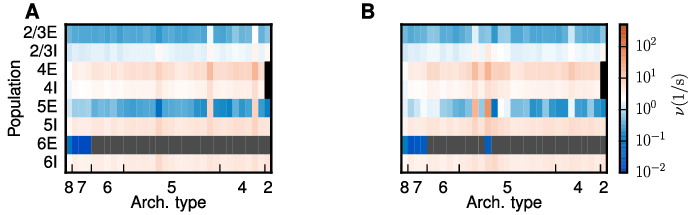}
\par\end{centering}
\caption{\textbf{Unstable fixed points in subsequent iterations}. Population
firing rates at the unstable fixed point as predicted by the mean-field
theory encoded in color for iterations 1 (\textbf{A}) and 2 (\textbf{B}).
Same display as in \prettyref{fig:bistability-multiarea}.}
\label{fig:unstable_FP}
\end{figure}

To control the separatrix we need to find the unstable fixed point
of the system. This is nontrivial since the numerical integration
of \prettyref{eq:int_siegert} for finding equilibria by construction
only converges to stable fixed points. If the unstable fixed point
has only one repelling direction (\prettyref{fig:eigenspace}A), it
constitutes a stable attractor on the $N-1$ dimensional separatrix.
The separatrix is a stable manifold \citep{Strogatz94}, and therefore
a trajectory originating in its vicinity but not near an unstable
fixed point initially stays in the neighborhood. If an initial condition
just outside the separatrix is close to the basin of attraction of
a particular unstable fixed point, the trajectory initially approaches
the latter. Close to the fixed point the velocity is small. Ultimately
trajectories diverge from the separatrix in the fixed point's unstable
direction, as illustrated in \prettyref{fig:mf-multiarea}A. In conclusion,
we expect a local minimum in the velocity along the trajectories close
to the unstable fixed point. To estimate the location of the unstable
fixed point in this manner, we need to find initial conditions close
to the separatrix. Naively, we would just fix the value of $\kappa$
and vary the initial condition. However, due to the high dimensionality
of our system this is not feasible in practice. Instead, we vary $\kappa$
for a fixed initial condition. \prettyref{fig:mf-multiarea}B shows
the firing rate averaged across populations for two trajectories starting
close to the separatrix, where the first one converges to the LA fixed
point and the second one to the HA state. The trajectories diverge
near the unstable fixed point and thus we define the last local minimum
of the Euclidean norm of the velocity vector as the critical time
$s_{\mathrm{c}}$ at which we assume the system to be close to the
unstable fixed point (\prettyref{fig:mf-multiarea}C). We find four
relevant and distinct unstable fixed points, of which two are shown
in \prettyref{fig:unstable_FP}.

To counteract the shift of the separatrix caused by the increase in
$\kappa$, we follow the procedure described in \nameref{sec:Theory-and-Methods}.
We subject the modifications of connectivity to the additional following
constraints. In line with the anatomical literature, we do not allow
for changes of the connectivity that would lead to cortico-cortical
connections originating in the granular layer 4 \citep{Felleman91_1},
and we also disallow inhibitory cortico-cortical connections, as the
vast majority of long-range connections are known to be excitatory
\citep{Salin95_107,Tomioka07_526}. In addition, we naturally restrict
indegrees to positive values. We find that four iterations (numbered
by index $j$) corresponding to the four distinct unstable fixed points
suffice to preserve the basin of attraction of the LA state with respect
to an increase of the external drive up to $\kappa=1.15$. In the
following we concentrate on iterations 1 and 2, where the second one
is also representative for iterations 3 and 4, which are qualitatively
alike. To derive the required modifications of the indegree matrix,
we decompose $\boldsymbol{K}$ into its $N$ eigenmodes and quantify
the contribution of each eigenmode to the shift of the unstable fixed
point (see \nameref{sec:Theory-and-Methods}). This allows us to identify
the most effective eigendirection: in each iteration $j$ there is
exactly one unstable eigendirection with an eigenvalue $\mathrm{Re}\left(\lambda_{\mathrm{c}}^{(j)}\right)>1$
(\prettyref{fig:eigenspace}A). The associated critical eigenvector
is approximately anti-parallel to the shift of the fixed point, $\boldsymbol{\delta\nu^{\ast}}$
(inset of \prettyref{fig:eigenspace}B), and of similar length. The
critical eigendirection (red dot in \prettyref{fig:eigenspace}B)
constitutes the most effective modification, giving the largest contribution
to the desired shift while requiring only a small change of $2.3\,\%$
in average total indegrees. In the chosen space of eigenmodes, the
modifications are minimal in the sense that only this most effective
eigenmode is changed.

\begin{figure}
\begin{centering}
\includegraphics{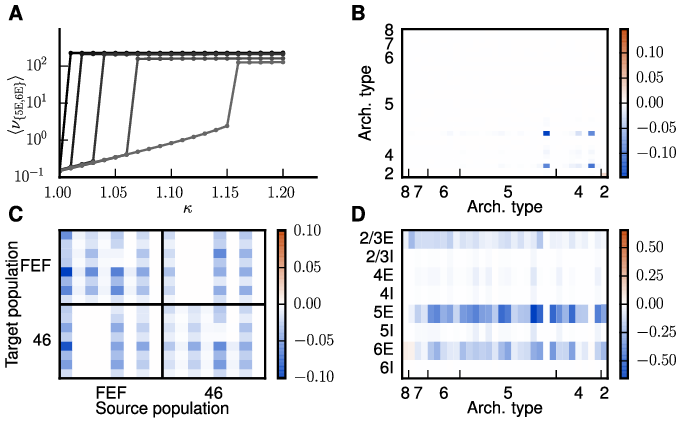}
\par\end{centering}
\centering{}\caption{\textbf{Altered phase space and modified connections. A} Firing rates
averaged across populations 5E and 6E and across areas for different
stages from the original model (black) to iteration 4 (light gray)
as a function of $\kappa$, predicted by the mean-field theory. \textbf{B}
Relative changes in the indegree $\delta K_{AB}/\sum_{B}K_{AB}$ between
areas $A,B$ in the first iteration. \textbf{C} Layer-specific relative
changes $\delta K_{ln}/\sum_{n}K_{ln}$ in the connections within
and between areas FEF and 46, for the first iteration. Populations
are ordered from 2/3E (left) to 6I (right) on the horizontal axis
and from 6I (bottom) to 2/3E (top) on the vertical axis as in panel
D. \textbf{D} Relative changes in population-specific indegrees summed
over target populations, $\sum_{l}\delta K_{ln}/\sum_{l}K_{ln}$,
combined for iterations two, three and four.}
\label{fig:stabilization_multiarea}
\end{figure}

The associated eigenvector $u_{\mathrm{c}}^{(1)}$ predominately points
into the direction of populations 4E and 5E of areas FEF and 46 (\prettyref{fig:eigenspace}C),
while $u_{\mathrm{c}}^{(2)}$ has large entries in the 5E populations
of two areas (\prettyref{fig:eigenspace}D). The high rates of these
populations at the unstable fixed points (cf. \prettyref{fig:unstable_FP}A,B
with \prettyref{fig:eigenspace}C,D) reflect that the instability
is caused by increased rates in excitatory populations, particularly
in population 5E. Each iteration shifts the transition to the HA state
(the value of $\kappa$ for which the separatrix crosses the initial
condition) to higher values of $\kappa$ and increases the attainable
rates of populations 5E and 6E in the LA state (\prettyref{fig:stabilization_multiarea}A).
After all four iterations, the average total indegrees (summed over
source populations) of the system are changed by $11.3\%$. The first
iteration mainly affects connections within and between areas 46 and
FEF (\prettyref{fig:stabilization_multiarea}B). In particular, the
excitatory loops between the two areas are reduced in strength, especially
those involving layer 5 (\prettyref{fig:stabilization_multiarea}C).
We thus identify two areas forming a critical loop. In the remaining
iterations, the changes are spread across areas and especially connections
originating in layer 5 are weakened (\prettyref{fig:stabilization_multiarea}D).
In conclusion, the method identifies critical structures in the model
both on the level of areas and on the level of layers and populations,
and leads to a small but specific structural change of the model.

\subsection{Analysis of the modifications}

\begin{figure}
\begin{centering}
\includegraphics{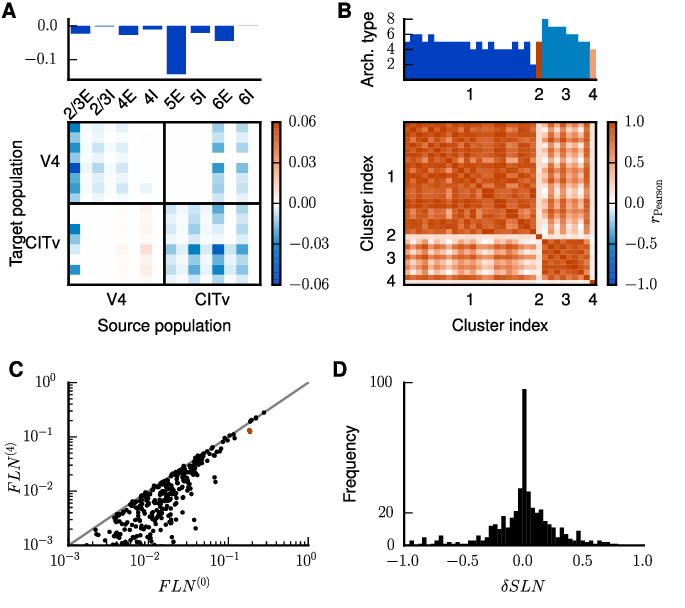}
\par\end{centering}
\centering{}\caption{\textbf{Analysis of changes in connectivity. A }Top panel: relative
changes in population-specific intrinsic indegrees summed over target
populations and averaged across areas, $\langle\sum_{i}\delta K_{AA}^{ij}/\sum_{i}K_{AA}^{ij}\rangle_{A}$.
Bottom panel: changes in the indegrees within and between exemplary
areas V4 and CITv relative to the total indegrees of the target populations,
i.e., $\delta K_{AA}^{ij}/\sum_{j}K_{AA}^{ij}$. Populations ordered
as in \prettyref{fig:stabilization_multiarea}C. \textbf{B} Pearson
correlation coefficient of the changes of the internal indegrees $\delta K_{AA}^{ij}$
between all pairs of the 32 areas. Areas ordered according to hierarchical
clustering using a farthest point algorithm \citep{Voorhees1986}.
The heights of the bars on top of the matrix indicate the architectural
types of the areas (types $1$ and $3$ do not appear in the model)
with color representing the respective clusters. \textbf{C} $FLN$
of the modified connectivity after 4 iterations versus the original
$FLN$ of the model. Only $FLN>10^{-3}$ are shown for a better overview.
The overlapping red dots represent the connections between areas 46
and FEF. Unity line shown in gray.\textbf{ D }Histogram of the cumulative
changes in $SLN$ over all four iterations ($\delta SLN=SLN^{(4)}-SLN^{(0)}$).}
\textbf{\label{fig:conn_analysis}}
\end{figure}

In the following we analyze the modifications of the connectivity
with respect to the internal and inter-area connections in detail.
The intrinsic circuits of the areas are modified in different directions,
as shown for two exemplary areas V4 and CITv in \prettyref{fig:conn_analysis}A.
Despite this heterogeneity, significant changes affect mostly excitatory-excitatory
connections (\prettyref{fig:conn_analysis}A, bottom panel) with connections
from population 5E experiencing the most significant changes (top
panel of \prettyref{fig:conn_analysis}A). In fact, the anatomical
data \citep{Binzegger04} underlying the microcircuit model \citep{Potjans14_785}
contain only two reconstructed excitatory cells from layer 5, but
considerably more for other cell types, indicating a higher uncertainty
for layer 5 connections. \prettyref{fig:conn_analysis}B shows the
correlation between intrinsic connectivity changes for all pairs of
areas, with areas ordered according to a hierarchical clustering using
a farthest point algorithm \citep{Voorhees1986} on the correlation
matrix. We find four clusters each indicating a group of areas which
undergo changes with similar patterns. The groups are displayed in
different colors in the histogram in \prettyref{fig:conn_analysis}B.
The areas of the model are categorized into architectural types based
on cell densities and laminar thicknesses (see \nameref{sec:Theory-and-Methods}).
Areas with architectural type 4, 5 and 6 are distributed over several
clusters. We can interpret this as a differentiation of these types
into further subtypes. The resulting changes of the intra-areal connectivity
are small (\prettyref{fig:conn_analysis}A), but still significant
for network stability.

Connections between areas can be characterized by their $FLN$ and
$SLN$ (see \nameref{sec:Theory-and-Methods}). The $FLN$ reflects
the overall strength of an inter-areal connection and is only weakly
affected across connections (\prettyref{fig:conn_analysis}C), with
a correlation between original and modified logarithms of $FLN$ of
$r_{\mathrm{Pearson}}=0.79$. Significant variations in the $FLN$
occur mostly for very weak connections that are likely to have substantial
relative uncertainties in the experimental data. The two overlapping
red dots in \prettyref{fig:conn_analysis}C represent the connections
between areas 46 and FEF, which are modified in the first iteration
(\prettyref{fig:stabilization_multiarea}C). The $SLN$ determines
the laminar pattern of the location of source neurons for cortico-cortical
connections. Overall, data are available for $24\%$ of the inter-areal
connections in the parcellation of Felleman \& van Essen \citep{Felleman91_1},
while the $SLN$ for the rest are derived from the sigmoidal law.
The majority of connections undergo small changes in their laminar
source pattern (\prettyref{fig:conn_analysis}D) and connections with
large modifications ($\left|\delta SLN\right|>0.5$) are weak (average
$\overline{FLN}=6\cdot10^{-4}$ compared to $\overline{FLN}=10^{-2}$
in the model as a whole). Because weak connections are represented
by low counts of labeled neurons, they have a relatively large uncertainty
in their laminar patterns, justifying larger adjustments. Spearman's
rank correlation between the $SLN$ of the original model that were
directly taken from experiments and the logarithmic ratios of cell
densities is $\rho=-0.63$ ($p=3\cdot10^{-11}$, p-value of a two-sided
test for uncorrelated data). For the modified model, we take the $SLN$
of all connections into account and obtain $\rho=-0.40$ ($p=6\cdot10^{-20})$,
indicating a reduced, but still significant, monotonic dependence
between $SLN$ and the logarithmic ratios of cell densities.

To judge the size of the modification to the connectivity, we compare
it to the variability of measured cortico-cortical connection densities
\citep{Markov2014_17}. We quantify the latter as the average inter-individual
standard deviation of the logarithmic $FLN$, i.e., $\sigma=\left<\sqrt{\overline{\left(\log FLN-\overline{\log FLN}\right)^{2}}}\right>$,
where the overbar $\av{\dotv}$ denotes the average over injections
and $\langle\dotv\rangle$ the average over connections. This variability
equals $2.17$ while the average modification of the logarithmic $FLN$
is $1.34$. The main experimental connection probabilities used to
construct the intra-areal connectivity of the model have an average
relative standard deviation of $30\%$ across electrophysiological
experiments (cf. Table 1 of \citep{Potjans14_785}) while the intra-areal
connection probabilities of the model are modified by $9\%$ on average.
The authors of \citep{Scannel00} report even greater variability
in their review on cortico-cortical and thalamocortical connectivity.
These considerations show that on average, the changes applied to
the connectivity are well within the uncertainties of the data. Overall,
35 out of 603 connections were removed from the network. In the CoCoMac
database, $83\,\%$ of these are indicated by only a single tracer
injection, while the overall proportion of connections measured by
a single injection is $59\,\%$.

\begin{figure}
\begin{centering}
\includegraphics{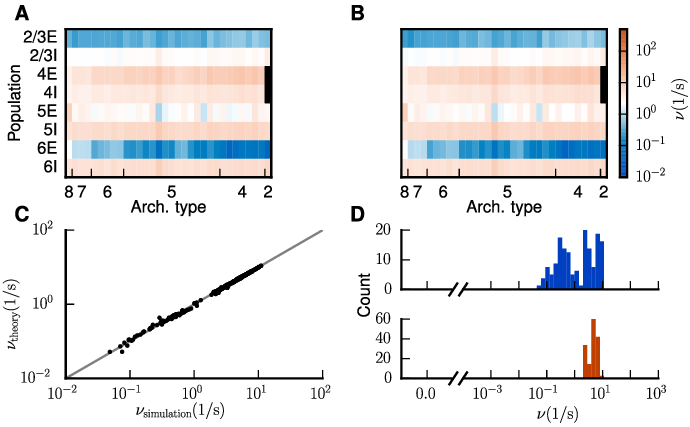}
\par\end{centering}
\begin{centering}
\par\end{centering}
\caption{\textbf{Improved low-activity fixed point of the model. }Population-averaged
firing rates for $\kappa=1.125$ encoded in color (\textbf{A}) predicted
by the analytical theory and (\textbf{B}) obtained from the full simulation
of the spiking network. Same display as in \prettyref{fig:bistability-multiarea}.
\textbf{C} Analytical versus simulated firing rates (black dots) and
identity line (gray). \textbf{D} Histogram of population-averaged
simulated firing rates. Same display as in \prettyref{fig:bistability-multiarea}.}
\label{fig:simulation}
\end{figure}

For the modified connectivity and $\kappa=1.125$, which we choose
to avoid being too close to the transition (\prettyref{fig:stabilization_multiarea}A),
the theory predicts average rates in populations 5E and 6E of $1.3$
and $0.18\spikess$, which is closer to experimentally observed rates
compared to the original model. Furthermore we find that the modified
connectivity allows us to decrease the inhibition in the network to
$g=-11$. Simulating the full spiking network model then results in
reasonable rates across populations and areas (\prettyref{fig:simulation}B,
D). The average rates in populations 5E and 6E are increased compared
to a simulation of the original model from $0.09$ and $2\cdot10^{-5}\spikess$
to $3.0$ and $0.4\spikess$, respectively. All populations exhibit
firing rates within a reasonable range of $0.05$ to $30\spikess$
(\prettyref{fig:simulation}D), as opposed to the original state in
which a considerable fraction of excitatory neurons is silent (\prettyref{fig:bistability-multiarea}E).
The theoretical prediction is in excellent agreement with the rates
obtained in the simulation (\prettyref{fig:simulation}A, C). Small
discrepancies are caused by violations of the employed assumptions,
i.e., Poissonian spiking statistics \citep{Fourcaud02}. Differences
between theory and simulation are small, and negligible for the central
aim of the study: the integration of activity constraints into the
data-driven construction of multi-scale neuronal networks.

\newpage

\section{Discussion\label{sec:Discussion}}

This study investigates the link between experimentally measured
structural connectivity and neuronal activity in a multi-scale spiking
network model of the vision-related areas of macaque cortex \citep{Schmidt16_arxiv_v4}.
We devise a theoretical method that systematically combines anatomical
connectivity with physiological activity constraints. Already weak
constraints, demanding the activity to neither vanish nor be pathologically
high, yield a set of specific but small structural modifications necessary
to increase the model's excitation to a realistic level. We do not
fit the model parameters to experimental activity data in the sense
of minimizing an error function, since the sparse relevant experimental
data do not allow for defining such a function in a meaningful way.
Nevertheless, we considerably improve the network activity with a
change in only a small set of parameters. The procedure constrains
the experimentally obtained connectivity maps to a realization that
is compatible with physiological experiments. This establishes a path
from experimentally observed activity to specific hypotheses about
the anatomy and demands for further experiments.

Connections are modified both inside and between cortical areas, on
average well within the known uncertainties of the underlying data.
The model areas are based on the early sensory cortex model presented
in \citep{Potjans14_785}. This circuit is adapted to individual areas
by taking into account neuronal densities and laminar thicknesses.
The model definition renders areas with equal architectural type similar
in their internal connectivity, a drastic but inevitable simplification
due to the lack of more detailed experimental data. The proposed method
softens this assumption: small adaptations of internal connectivity
distinguish the architectural types into further subtypes. These modifications
are significant for the global stability of the network. Thus, our
approach enables purely anatomy-based area categorizations to be refined
with dynamical information.

Connections between areas are changed in terms of total strength and
laminar patterns. Overall, the changes are small, but significant
for specific connections. The loop formed by areas 46 and FEF is critical
to the global stability of the network. Both areas have been investigated
in \citep{Markov2014_17}, albeit in a different parcellation scheme
than the scheme used here \citep{Felleman91_1}. Our method suggests
a weaker coupling of these two areas than found in the anatomical
data set. Uncertainties, partly due to the mapping between parcellations,
leave room for this interpretation. Areas 46 and FEF belong to prefrontal
cortex and are multimodal, indicating that the influence of other
parts of cortex could stabilize them, a mechanism outside the scope
of the present model of vision-related areas. Both explanations can
be tested either in an experimental study or with an extended model.

A few weak connections undergo large changes in their laminar patterns.
With the present activity constraints, the method hereby weakens hierarchical
relations in the model structure, indicating that preserving these
relations requires additional dynamical constraints such as layer-specific
coherence between areas \citep{Kerkoerle14_14332,Bastos15_390}. Conversely,
it is possible to achieve satisfactory population rates with a less
pronounced hierarchical structure. 

Our analysis reveals that layer 5 excitatory cells play a critical
role in the model's dynamics, in line with the observed ongoing activity
in mammalian neocortex \citep{Sanchez-Vives00_1027,beltramo2013_227}.
This critical role is often attributed to single-neuron properties,
with a subset of layer 5 neurons displaying pacemaker activity \citep{Bon-Jego2007,Lorincz15_5442,Neske2015_1089}.
In addition, we here find that the network architecture itself already
explains the strong impact of layer 5 on the phase space of the network,
suggesting that single-neuron properties and network structure jointly
enable layer 5 to exert its dominant influence.

The cortico-cortical connectivity of the model is compiled from the
extensive dataset of \citep{Markov2014_17} combined with the CoCoMac
database \citep{Stephan01_1159,Bakker12_30}, which collects data
from hundreds of tracing studies. One could consider alternative
methods for combining this information into one connectivity graph,
for instance taking into account how consistently a given connection
is reported across studies \citep{Schmitt14_1}, and compare different
methods by analyzing the resulting network dynamics. The presented
mean-field theory could then be used to estimate the firing rates
of each network instance without performing time-consuming simulations.
However, here we first choose the integrative approach that accumulates
all available experimental evidence into a single model and afterwards
modify the resulting connectivity with our analytical procedure, thereby
effectively discarding uncertain connections.

We restrict this study to networks of leaky integrate-and-fire model
neurons, consistent with the key concept of the models we consider:
individual cells are modeled in a simple manner to expose the impact
of structural connectivity on the network dynamics. Moreover, the
current-based leaky integrate-and-fire neuron can reproduce in-vivo
like activity \citep{Rauch03,Jolivet06} and is analytically tractable,
which enables the identification of mechanisms underlying specific
network effects. More complex neuron models can be incorporated into
the method by replacing the gain function of each neuronal population
with an analytical expression or an interpolated function obtained
from spiking single-neuron simulations. For example, one could use
a conductance-based point-neuron model for which the network dynamics
can be described by population rate models \citep{Shriki03}, featuring
a non-monotonic gain function: the gain is reduced if excitatory and
inhibitory inputs are increased in a balanced manner \citep{Kuhn04}.
Generally, this renders a system more stable. However, the bistability
considered in our work is caused by excitatory inputs. Since conductance-based
models also have a monotonically increasing gain function in dependence
of the excitatory conductance alone, we expect the bistability to
occur for such models as well. 

Importantly, our method only requires a description of the system's
fixed points and their dependence on model parameters. The employed
theory uses the diffusion approximation to derive a self-consistency
equation for the stationary population rates valid for high indegrees
and small synaptic efficacies. These requirements are fulfilled in
the multi-area model and therefore the theoretical prediction agrees
with the stationary activity of the simulation. Moreover, the theory
predicts the bistability of the model, which is non-trivial, as the
mean-field assumption of Poisson statistics of the activity is generally
violated in the high-activity state. Nevertheless, since the activity
in this state is mostly driven by strong mean inputs and the theory
converges to the noiseless solution in the mean-driven limit, its
predictions still provide viable approximations. The firing rates
in the unstable fixed points are predominantly low, while the exceptions
with very high rates are again mean-dominated. Presumably this explains
the accuracy by which the theory predicts the locations of these fixed
points and the resulting global stability properties.

Since the high-activity attractor of the model under consideration
is unrealistic, we aim to prevent a transition to this state. However,
in the high-dimensional system it is not a priori clear in which direction
the separatrix has to be shifted to ensure stable dynamics. We therefore
choose a pragmatic approach and shift the separatrix back to its initial
location, inverting the shift which reduced the global stability of
the low-activity fixed point. We achieve this to a good approximation
by preserving the location of unstable fixed points on the separatrix.
To this end, we use a linearization around these locations and an
eigenmode decomposition to identify the set of connections to adjust.
In the multi-area model, this linearization is justified because the
system operates close to an instability so that only minor modifications
are required. The method can be generalized to larger modifications
by changing parameters iteratively in small steps.

Biological networks have various stabilization mechanisms not considered
here, which render them less critical. For instance, during growth,
homeostatic mechanisms guide the system toward the right structure.
Furthermore, short-term synaptic plasticity (reviewed in \citep{Morrison08_459}),
homeostatic synaptic scaling \citep{Turrigiano98} and spike-frequency
adaptation (e.g. summarized in \citep{Benda03_2523}) may prevent
the system from entering the high-activity state. However, introducing
these self-organizing mechanisms increases model complexity, causing
a more intricate relation between structure and activity. Therefore,
we start from a mean-field description on the level of neuronal populations,
ignoring details of synaptic dynamics. Mild constraints on the activity
lead to a network structure within the anatomical range of parameters.
This network yields globally stable activity, suggesting that additional
stabilization mechanisms are not required to achieve this. Nonetheless,
they can potentially render the network more robust against external
stimulation.

The observed inter-individual variability may reflect that mechanisms
of self-organization and homeostasis find structurally different implementations
of the same function. Thus, in studies across individuals we cannot
expect that progress in experimental technology narrows down the variability
of parameters indefinitely. The combined ranges of parameters rather
specify the solution space, and our method provides a way to find
a particular solution. 

In principle, the method applies to any model parameter. It would
be possible to modify, for example, synaptic weights. Since experimental
data on synaptic weights are sparse, this is another natural choice.
Moreover, such an analysis may provide hints about suitable synaptic
plasticity rules that dynamically stabilize the model. The method
can be applied to networks with more complex sets of attractors compared
to the bistable case considered here. Though in high-dimensional systems
such as our multi-area model, a larger number of attractors would
make it more challenging to find all relevant unstable fixed points,
the underlying idea of preserving the location of a separatrix is
general. In contrast to the model considered here, transitions between
fixed points can have a functional meaning in certain neuronal networks
with multiple attractors. The specific location of the separatrix
is then functionally relevant. Our method exposes the sensitivity
of the location of the separatrix to certain model parameters and
allows controlling its location in a specific manner.

In this work, we analyzed the global stability properties of the neuronal
network on the population level. In contrast, Ostojic \citep{Ostojic14}
performs a local stability analysis on the level of single neurons
of an initially stable fixed point in a system with only one attractor.
The author investigates the point at which the real parts of the eigenvalues
of the Jacobian matrix evaluated at this fixed point become positive,
i.e., the fixed point turns spectrally unstable and the system undergoes
a transition to a heterogeneous asynchronous state. Analyzing the
spectral stablilty on the single-neuron level does not reveal the
global stability properties required in the current work: While a
local stability analysis only considers infinitesimal perturbations,
studying the basin of attraction gives information about the size
of fluctuations against which the fixed point is stable. We expect
both attractors to be spectrally stable because they do not show strong
rate fluctuations and the mean-field theory predicts the activity
accurately in both cases. Nevertheless, a heterogeneous state could
occur if the synaptic weights were increased or the external drive
was stronger. However, \citep{Goedeke16_arxiv} show that the transition
in stochastic systems that quantitatively resemble the spiking network
\citep{Grytskyy13_131}, does not coincide with the loss of spectral
stability.

One striking feature of the heterogeneous state is bursty spiking
behavior of individual cells. Bursty spiking is also observed in the
multi-area model for increased synaptic weights of inter-area connections
\citep{Schmidt16_arxiv_v4}. The fixed points cannot be accurately
described in this case because the fluctuations need to be taken into
account \citep{Mastroguiseppe16_arxiv}. Simulations show (figures
4 and 5 of \citep{Schmidt16_arxiv_v4}) that the modifications obtained
in this study are still able to prevent the system from a transition
to a HA attractor also in the presence of bursting neurons. This indicates
that the phase space does not change qualitatively and our results
are robust against such bursting behavior.

Experimental data on stationary activity in cortex are sparse. We
therefore restrict ourselves to fundamental constraints and increase
the drive to the model in an area-unspecific way to fulfill them.
The resulting heterogeneity of the firing rates across areas is thus
not imposed by the method, but rather arises from the connectivity
that remains strongly informed by anatomical data. Alternatively,
one could predefine a desired state and investigate the parameter
changes necessary to achieve it.

The presented analytical method that combines anatomy and activity
data into a consistent model is restricted to stationary firing rates.
In future studies, also higher-order statistical measures of activity
can be used as constraints. Resting-state fMRI, for example, provides
information on the functional connectivity between areas as a second-order
measure. When combined with analytical predictions of functional connectivity,
the method may shed light on the anatomical connection patterns underlying
inter-area communication.

\section{Methods\label{sec:Theory-and-Methods}}

In this study, we model single cells as leaky integrate-and-fire model
neurons with exponentially decaying postsynaptic currents. \prettyref{tab:Parameters}
specifies the model parameters. 
\begin{table}[h]
\begin{centering}
\caption{Specification of the neuron and synapse parameters.\label{tab:Parameters}}
\begin{tabular}{@{\hspace*{1mm}}p{1.4cm}@{}||@{\hspace*{1mm}}p{1.4cm}@{}@{\hspace*{1mm}}p{1.1cm}@{}||@{\hspace*{1mm}}p{1.1cm}@{}||@{\hspace*{1mm}}p{1.1cm}@{}@{\hspace*{1mm}}p{1.1cm}||@{\hspace*{1mm}}p{1.1cm}||@{\hspace*{1mm}}p{1.1cm}||@{\hspace*{1mm}}p{1.1cm}||@{\hspace*{1mm}}p{1.1cm}||@{\hspace*{1mm}}p{1.1cm}}
\hline 
\multicolumn{11}{>{\columncolor{parametergray}}c}{\textbf{Synapse parameters}}\tabularnewline
\multicolumn{2}{@{\hspace*{1mm}}p{2.7cm}@{}}{\textbf{Name}} & \multicolumn{3}{@{\hspace*{1mm}}p{3.65cm}@{}}{\textbf{Value}} & \multicolumn{6}{@{\hspace*{1mm}}p{6.6cm}}{\textbf{Description}}\tabularnewline
\hline 
\multicolumn{2}{>{\raggedright}p{1.35cm}}{$J\pm\delta J$} & \multicolumn{3}{>{\raggedright}p{4.3cm}}{$87.8\pm8.8\pA$} & \multicolumn{6}{>{\raggedright}p{5.1cm}}{excitatory synaptic strength}\tabularnewline
\multicolumn{2}{l}{$g$} & \multicolumn{3}{@{\hspace*{1mm}}p{3.3cm}@{}}{$-16$ (\prettyref{fig:bistability-multiarea})

$-11$ (\prettyref{fig:simulation})} & \multicolumn{6}{@{\hspace*{1mm}}p{6.6cm}@{}}{relative inhibitory synaptic strength}\tabularnewline
\multicolumn{2}{l}{$d_{\mathrm{e}}\pm\delta d_{\mathrm{e}}$} & \multicolumn{3}{@{\hspace*{1mm}}p{3.3cm}@{}}{$1.5\pm0.75\ms$} & \multicolumn{6}{@{\hspace*{1mm}}p{6.6cm}@{}}{local excitatory transmission delay}\tabularnewline
\multicolumn{2}{l}{$d_{\mathrm{i}}\pm\delta d_{\mathrm{i}}$} & \multicolumn{3}{@{\hspace*{1mm}}p{3.3cm}@{}}{$0.75\pm0.375\ms$} & \multicolumn{6}{@{\hspace*{1mm}}p{6.6cm}@{}}{local inhibitory transmission delay}\tabularnewline
\multicolumn{2}{l}{$d\pm\delta d$} & \multicolumn{3}{@{\hspace*{1mm}}p{3.3cm}@{}}{$d=s/v_{\mathrm{t}}\pm\frac{1}{2}s/v_{\mathrm{t}}$} & \multicolumn{6}{@{\hspace*{1mm}}p{6.6cm}@{}}{inter-areal transmission delay, with $s$ the distance between areas}\tabularnewline
\multicolumn{2}{l}{$v_{\mathrm{t}}$} & \multicolumn{3}{@{\hspace*{1mm}}p{3.3cm}@{}}{$3.5\,\mathrm{m/s}$} & \multicolumn{6}{@{\hspace*{1mm}}p{6.6cm}@{}}{transmission speed}\tabularnewline
\end{tabular}
\par\end{centering}
\centering{}%
\begin{tabular}{@{\hspace*{1mm}}p{2.7cm}@{}@{\hspace*{1mm}}p{3.65cm}@{}@{\hspace*{1mm}}p{6.6cm}}
\hline 
\multicolumn{3}{>{\columncolor{parametergray}}c}{\textbf{Neuron model}}\tabularnewline
\textbf{Name} & \textbf{Value } & \textbf{Description}\tabularnewline
\hline 
$\taum$ & $10\ms$ & membrane time constant\tabularnewline
$\taur$ & $2\ms$ & absolute refractory period\tabularnewline
$\taus$ & $0.5\ms$  & postsynaptic current time constant\tabularnewline
$C_{\mathrm{m}}$ & $250\pF$  & membrane capacity\tabularnewline
$V_{\mathrm{r}}$ & $-65\mV$ & reset potential\tabularnewline
$\theta$ & $-50\mV$ & fixed firing threshold\tabularnewline
$\EL$ & $-65\mV$ & leak potential\tabularnewline
\end{tabular}
\end{table}
In the diffusion approximation, the dynamics of the membrane potential
$V$ and synaptic current $I_{\mathrm{s}}$ is \citep{Fourcaud02}

\begin{eqnarray}
\taum\frac{\d V}{\d t} & = & -V+I_{\mathrm{s}}(t)\nonumber \\
\taus\frac{\d I_{\mathrm{s}}}{\d t} & = & -I_{\mathrm{s}}+\mu+\sigma\sqrt{\taum}\xi(t),\label{eq:diffeq_iaf-1-1}
\end{eqnarray}
where $\tauM$ is the membrane time constant and $\tau_{\mathrm{s}}$
the synaptic time constant, respectively. The membrane resistance
$\taum/C_{\mathrm{m}}$ has been absorbed into the definition of the
current. The input spike trains are approximated by a white noise
current with fluctuations $\propto\sigma^{2}$ and mean value $\mu$.
Here $\xi$ is a centered Gaussian white process satisfying $\langle\xi(t)\rangle=0$
and $\langle\xi(t)\xi(t^{\prime})\rangle=\delta(t-t^{\prime})$. Whenever
the membrane potential $V$ crosses the threshold $\theta,$ the neuron
emits a spike and $V$ is reset to the potential $V_{\mathrm{r}}$,
where it is clamped during $\taur$. All neurons in one population
have identical parameters, so that we can describe the network activity
in terms of population-averaged firing rates $\nu_{i}$ that depend
on population-dependent input $\mu_{i},\sigma_{i}$ determined by
the connectivity. Using the Fokker-Planck formalism, the stationary
firing rates for each population $i$ are given by \citep{Fourcaud02}
\begin{eqnarray}
\frac{1}{\nu_{i}} & = & \taur+\taum\sqrt{\pi}\int_{\frac{V_{r}-\mu_{i}(\boldsymbol{A})}{\sigma_{i}(\boldsymbol{A})}+\gamma\sqrt{\frac{\taus}{\taum}}}^{\frac{\theta-\mu_{i}(\boldsymbol{A})}{\sigma_{i}(\boldsymbol{A})}+\gamma\sqrt{\frac{\taus}{\taum}}}e^{x^{2}}\ (1+\erf(x))\,\d x\nonumber \\
 & \eqqcolon & 1/\Phi_{i}(\boldsymbol{\nu},\boldsymbol{A})\label{eq:siegert_2D}\\
\mu_{i}(\boldsymbol{A}) & = & \taum\sum_{j}K_{ij}J{}_{ij}\nu_{j}+\taum K_{\mathrm{ext}}J_{\mathrm{ext}}\nuext\label{eq:mean}\\
\sigma_{i}^{2}(\boldsymbol{A}) & = & \taum\sum_{j}K_{ij}J_{ij}^{2}\nu_{j}+\taum K_{\mathrm{ext}}J_{\mathrm{ext}}^{2}\nuext,\label{eq:sigma}
\end{eqnarray}
which is correct up to linear order in $\sqrt{\taus/\taum}$ and where
$\gamma=|\zeta(1/2)|/\sqrt{2},$ with $\zeta$ denoting the Riemann
zeta function \citep{Abramowitz74}. Here, $\boldsymbol{A}$ is chosen
from the set of model parameters $\{\boldsymbol{K},\boldsymbol{J},\nuext,\dots\}$.
If $\boldsymbol{A}$ is a matrix, we vectorize it by concatenating
its rows and indicate this by lower case, i.e., $\boldsymbol{a=}\left(a_{00},a{}_{01},\dots,a{}_{0N},a{}_{10},\dots,a{}_{1N},a_{N0}\dots,a{}_{NN}\right)=\mathrm{vec}(\boldsymbol{A}^{\mathrm{T}})^{\mathrm{T}}$
following \citep{magnus95}. If the chosen parameter is a scalar
we denote it with $a$.

We find the fixed points of \prettyref{eq:siegert_2D} by solving
the first-order differential equation \prettyref{eq:int_siegert}
\citep{Wong_06}

\[
\dot{\boldsymbol{\nu}}\coloneqq\frac{\d\boldsymbol{\nu}}{\d s}=\boldsymbol{\Phi}(\boldsymbol{\nu},\boldsymbol{A})-\boldsymbol{\nu},
\]
using the classical fourth-order Runge-Kutta method (RK4) with step
size $h=0.01$, where $s$ denotes a dimensionless pseudo-time. The
same approach can be used to solve the activity on a single neuron
level \citep{Sadeh15}. Note that \prettyref{eq:int_siegert} does
not reflect the real time evolution of the population rates, but rather
is a mathematical method to obtain the system's  fixed points. In
contrast to \citep{Wong_06} we do not only search for stable fixed
points, but also use \prettyref{eq:int_siegert} to obtain unstable
attractors (cf. \nameref{sec:Results}), an idea originating from
the study of simple attractor networks (\citep{Amit97} esp. their
figure 2 and eq. 12).

In a bistable situation, the initial condition of \prettyref{eq:int_siegert}
determines which fixed point the system settles in. However, studying
the behavior for a particular initial condition is of minor interest,
since the actual spiking network is a stochastic system which fluctuates
around the fixed points of the deterministic system defined by \prettyref{eq:int_siegert}.
Even if we knew that \prettyref{eq:int_siegert} would relax to the
LA fixed point for one particular initial condition, this would not
necessarily imply that this state is indefinitely stable. Global stability
is determined by the size of the basin of attraction of the LA fixed
point.

In the following, we derive the equations leading us to targeted modifications
of a parameter $\boldsymbol{b}$ necessary to compensate for the changes
in the global stability induced by the change of another parameter
$\boldsymbol{a}$. To this end, we study the behavior of the fixed
points with respect to an infinitesimal change $\boldsymbol{\delta a}=\boldsymbol{a^{\prime}}-\boldsymbol{a}$
in the chosen model parameter. Let $\boldsymbol{\nu^{\ast}}(\boldsymbol{a})$
and $\boldsymbol{\nu^{\ast}}(\boldsymbol{a^{\prime}})$ be the corresponding
locations of the fixed points and $\boldsymbol{\delta\nu^{\ast}}=\boldsymbol{\nu^{\ast}}(\boldsymbol{a^{\prime}})-\boldsymbol{\nu^{\ast}}(\boldsymbol{a})$
their separation. We can then expand $\boldsymbol{\nu^{\ast}}(\boldsymbol{a^{\prime}})$
into a Taylor series up to first order in $\boldsymbol{\delta a}$
and obtain
\begin{eqnarray}
\boldsymbol{\nu^{\ast}}(\boldsymbol{a^{\prime}}) & = & \boldsymbol{\nu^{\ast}}(\boldsymbol{a})+\boldsymbol{\Delta_{a}\delta a}\nonumber \\
\Leftrightarrow\boldsymbol{\delta\nu^{\ast}} & = & \boldsymbol{\Delta_{a}}\boldsymbol{\delta a}\,,\label{eq:shift}
\end{eqnarray}
with
\begin{eqnarray}
\Delta_{a,ij} & = & \frac{\d\nu_{i}\left(\mu_{i},\sigma_{i}\right)}{\d a_{j}}\nonumber \\
\mbox{} & = & \frac{\d\Phi_{i}\left(\mu_{i},\sigma_{i}\right)}{\d a_{j}}\nonumber \\
 & = & \underbrace{\frac{\partial\Phi_{i}}{\partial\mu_{i}}}_{S_{i}}\frac{\d\mu_{i}}{\d a_{j}}+\underbrace{\frac{\partial\Phi_{i}}{\partial\sigma_{i}}\frac{1}{2\sigma_{i}}}_{T_{i}}\frac{\d\sigma_{i}^{2}}{\d a_{j}},\label{eq:delta_ij-1}
\end{eqnarray}
where we notice that $S_{i}$ and $T_{i}$ only depend on the target
population $i$. We accordingly define two diagonal matrices $\boldsymbol{S}$
and $\boldsymbol{T}$ with $S_{ii}=S_{i}$ and $T_{ii}=T_{i}$. We
further define the connectivity matrix $\W=\boldsymbol{K}\circledast\J$,
where $\circledast$ denotes element-wise multiplication, also called
the Hadamard product (see \citep{cichocki09}, for a consistent set
of symbols for operations on matrices). The derivatives with respect
to $a_{j}$ have the compact expressions 
\begin{eqnarray*}
\frac{\d\mu_{i}}{\d a_{j}} & = & \frac{\partial\mu_{i}}{\partial a_{j}}+\sum_{n}\frac{\partial\mu_{i}}{\partial\nu_{n}}\frac{\d\nu_{n}}{\d a_{j}}\\
 & = & \left(\D_{\boldsymbol{a}}\boldsymbol{\mu}\right)_{ij}+\taum\sum_{n}(\underbrace{K_{in}J_{in}}_{=W_{in}})\Delta_{a,nj}\,,
\end{eqnarray*}
with the Jacobian $\left(\D_{\boldsymbol{a}}\mathbf{f}\right)_{ij}\coloneqq\frac{\partial f_{i}}{\partial a_{j}}$
of some vector-valued function $\mathbf{f}$ and
\[
\frac{\mathrm{d\sigma_{i}^{2}}}{\d a_{j}}=\left(\D_{\boldsymbol{a}}\boldsymbol{\sigma}^{2}\right)_{ij}+\taum\sum_{n}(\underbrace{K_{in}J_{in}^{2}}_{=W_{2,in}})\Delta_{a,nj}\,,
\]
where we use the Hadamard product again to define the matrix $\mathbf{W_{2}}\coloneqq\mathbf{K}\circledast\mathbf{J}\circledast\mathbf{J}$.
Inserting the total derivatives into \prettyref{eq:delta_ij-1}, we
derive the final expression for $\boldsymbol{\Delta_{a}}$, reading
\begin{eqnarray}
\boldsymbol{\Delta_{a}} & = & \boldsymbol{S}\left[\D_{\boldsymbol{a}}\boldsymbol{\mu}+\taum\boldsymbol{W}\boldsymbol{\Delta_{a}}\right]+\boldsymbol{T}\left[\D_{\boldsymbol{a}}\boldsymbol{\sigma}^{2}+\taum\boldsymbol{W_{2}}\boldsymbol{\Delta_{a}}\right]\nonumber \\
\Leftrightarrow\boldsymbol{\Delta_{a}} & = & \left[\unity-\underbrace{\taum(\boldsymbol{S}\boldsymbol{W}+\boldsymbol{T}\boldsymbol{W_{2})}}_{\eqqcolon\boldsymbol{M}}\right]^{-1}\underbrace{\left[\boldsymbol{S}\D_{\boldsymbol{a}}\boldsymbol{\mu}+\boldsymbol{T}\D_{\boldsymbol{a}}\boldsymbol{\sigma}^{2}\right]}_{\eqqcolon\boldsymbol{\bar{\Delta}_{a}}}\,,\label{eq:Delta_alpha}
\end{eqnarray}
where we use $\mathds{1}$ for the identity matrix and define the
effective connectivity matrix $\boldsymbol{M}$ and the matrix $\boldsymbol{\bar{\Delta}_{a}}$.
The latter has dimensionality $N\times P$, where $P$ is the dimension
of $\boldsymbol{a}$ (for example, $P=N^{2}$ for $\boldsymbol{a}=\boldsymbol{k}$
the vector of indegrees). With the aid of \prettyref{eq:Delta_alpha},
evaluating \prettyref{eq:shift} at the unstable fixed point predicts
the shift of the separatrix (\prettyref{fig:Brunel-2D}D) to linear
order. We now consider an additional parameter $\boldsymbol{b}$ which
is modified to counteract the shift of the unstable fixed point caused
by the change in parameter $\boldsymbol{a}$, i.e.,
\begin{eqnarray}
\boldsymbol{\Delta_{a}}\boldsymbol{\delta a} & \stackrel{!}{=} & -\boldsymbol{\Delta_{b}}\boldsymbol{\delta b}\nonumber \\
\Leftrightarrow\boldsymbol{\bar{\Delta}_{a}}\boldsymbol{\delta a} & = & -\boldsymbol{\bar{\Delta}_{b}}\boldsymbol{\delta b},\label{eq:counteract}
\end{eqnarray}
where we used that the inverse of ${\unity-\boldsymbol{M}}$ appears
on both sides of the equation and hence drops out. Note that the tuple
$\left(\boldsymbol{a},\boldsymbol{b}\right)$ may represent any combination
of model parameters, for example external input, indegrees, synaptic
weights, etc. For a particular choice of $\left(\boldsymbol{a},\boldsymbol{b}\right)$
we solve \prettyref{eq:counteract} for $\boldsymbol{\delta b}$.
For the illustrative example shown in \prettyref{fig:Brunel-2D},
where $a=\nu_{\mathrm{ext}}$ and $b=K$, \prettyref{eq:counteract}
simplifies to
\[
\delta K=\frac{\bar{\Delta}_{\nuext}\delta\nuext}{\bar{\Delta}_{K}}=\frac{K_{\mathrm{ext}}\delta\nuext}{\nu},
\]
since $S$ and $T$ appearing in the respective $\bar{\Delta}'$s
cancel each other. 

To determine critical connections in a more complex model, we choose
$\boldsymbol{b}=\boldsymbol{k}$, i.e. the vector of indegrees, and
solve \prettyref{eq:counteract} with a decomposition into eigenmodes.
We can write the right-hand side as
\begin{equation}
-\left(\boldsymbol{\bar{\Delta}_{k}}\boldsymbol{\delta k}\right)_{i}=-\sum_{j}\taum\left(S_{i}J_{ij}+T_{i}J_{ij}^{2}\right)\nu_{j}\delta K_{ij}\,.\label{eq:right_side_counteract}
\end{equation}
The equation holds because $\mu_{i},\sigma_{i}$ are only affected
by connections to population $i$, and therefore their derivatives
$\partial\mu_{i}/\partial k_{l},\:\partial\sigma_{i}/\partial k_{l}$
and hence $\bar{\Delta}_{k,il}$, vanish for $l\notin\left[(i-1)N+1,iN\right]$.
We now make the ansatz
\begin{equation}
\delta K_{ij}=\sum_{l}\frac{\epsilon_{l}}{\tauM}\frac{\left(\uvec^{l}\boldsymbol{v}{}^{l\mathrm{T}}\right)_{ij}}{\left(\boldsymbol{S}\boldsymbol{J}+\boldsymbol{T}(\boldsymbol{J}\circledast\boldsymbol{J})\right){}_{ij}}\,,\label{eq:ansatz_delta_K}
\end{equation}
which decomposes the changes $\boldsymbol{\delta K}$ into eigenmodes
of the effective connectivity. The $\uvec^{l}$ and $\boldsymbol{v}^{l}$
are the $l$-th right and left eigenvectors of $\boldsymbol{M}$ as
defined in \prettyref{eq:Delta_alpha}, fulfilling the bi-orthogonality
$\boldsymbol{v}^{l\mathrm{T}}\uvec^{n}=\delta_{ln}$ and the completeness
relation $\sum_{l}\uvec^{l}\boldsymbol{v}^{l\mathrm{T}}=\unity.$
Inserting \prettyref{eq:right_side_counteract} with \prettyref{eq:ansatz_delta_K}
into \prettyref{eq:counteract} yields 
\begin{equation}
\boldsymbol{\bar{\Delta}_{a}}\boldsymbol{\delta a}=-\sum_{l}\epsilon_{l}\uvec^{l}\boldsymbol{v}^{l\mathrm{T}}\boldsymbol{\nu}.\label{eq:bar_a_delta_a}
\end{equation}
Thus we can solve for $\epsilon_{l}$ by multiplying from the left
with $\boldsymbol{v}^{n\mathrm{T}}$

\begin{eqnarray*}
\underbrace{\boldsymbol{v}^{n\mathrm{T}}\boldsymbol{\bar{\Delta}_{a}}\boldsymbol{\delta a}}_{\eqqcolon\hat{a}^{n}} & = & -\sum_{l}\epsilon_{l}\underbrace{\boldsymbol{v}^{n\mathrm{T}}\uvec^{l}}_{\delta_{nl}}\underbrace{\boldsymbol{v}^{l\mathrm{T}}\boldsymbol{\nu}}_{\eqqcolon\hat{\nu}^{l}}\\
\epsilon_{n} & = & -\frac{\hat{a}^{n}}{\hat{\nu}^{n}}\,.
\end{eqnarray*}
Our goal is to find a set of connections which dominate the size of
the basin of attraction of the LA fixed point. Inserting \prettyref{eq:bar_a_delta_a}
into \prettyref{eq:shift} leads to

\begin{eqnarray*}
\boldsymbol{\delta\nu^{\ast}} & = & \sum_{l}(\unity-\boldsymbol{M})^{-1}\epsilon_{l}\uvec^{l}\boldsymbol{v}^{l\mathrm{T}}\boldsymbol{\nu}\\
 & = & \sum_{l}\frac{\epsilon_{l}}{1-\lambda_{l}}\boldsymbol{u}^{l}\hat{\nu}^{l}\\
 & = & \sum_{l}\frac{-\hat{a}^{l}}{1-\lambda_{l}}\boldsymbol{u}^{l},
\end{eqnarray*}
where $\lambda_{l}$ are the eigenvalues of $\boldsymbol{M}$, which
are either real or complex conjugate pairs since $\boldsymbol{M}\in\mathbb{R}^{N\times N}$.
To determine the influence of each eigenmode on the shift of the fixed
point, we project the eigenvectors $\boldsymbol{u}^{l}$ onto the
fixed-point shift $\boldsymbol{\delta\nu^{\ast}}$ by multiplying
each side with $\boldsymbol{\delta\nu^{\ast}}\boldsymbol{\delta\nu^{\ast}}^{\mathrm{T}}$
and solve again for $\boldsymbol{\delta\nu^{\ast}}$ to obtain
\begin{equation}
\boldsymbol{\delta\nu^{\ast}}=\sum_{l}\underbrace{\frac{-\hat{a}^{l}}{1-\lambda_{l}}\,\frac{\boldsymbol{\delta\nu^{\ast}}^{\mathrm{T}}\boldsymbol{u}^{l}}{\boldsymbol{\delta\nu^{\ast}}^{\mathrm{T}}\boldsymbol{\delta\nu^{\ast}}}}_{\eqqcolon\tilde{\eta}_{l}}\boldsymbol{\delta\nu^{\ast}},\label{eq:individual_shifts}
\end{equation}
where we define the (possibly complex-valued) coefficients $\tilde{\eta}_{l}.$
We aim at a decomposition of $\boldsymbol{\delta\nu^{\ast}}$ into
real components. If $\mathrm{Im}(\lambda_{l})=0$, $\tilde{\eta}_{l}$
is real, so we can work directly with $\eta_{l}\coloneqq\tilde{\eta}_{l}$.
Complex eigenvalues $\mathrm{Im}(\lambda_{l})\neq0$ and corresponding
eigenvectors come in conjugate pairs, so in this case we combine the
corresponding coefficients $\eta_{l}\coloneqq\tilde{\eta}_{l}+\tilde{\eta}_{l}^{*}$,
to have all contributions $\eta_{l}\in\mathbb{R}$ and $\sum_{l}\eta_{l}=1$
by construction \prettyref{eq:individual_shifts}. Each $\eta_{l}$
quantifies how much of the total fixed-point shift can be attributed
to the $l$-th eigenmode, which allows identification of the most
effective eigendirection (see \nameref{sec:Results}), where we apply
the ansatz \prettyref{eq:ansatz_delta_K} to a multi-area, multi-layer
model of the vision-related areas of macaque cortex.

The spiking simulations of the network model were carried out on the
JUQUEEN supercomputer \citep{stephan2015juqueen}. All simulations
were performed with NEST version 2.8.0 \citep{Nest280} with optimizations
for the use on the supercomputer which will be included in a future
release. The simulations use a time step of $0.1\ms$ and exact integration
for the subthreshold dynamics of the leaky integrate-and-fire neuron
model (reviewed in \citep{Plesser09_353}).

\subsection{Multi-area model}

The multi-area model of the vision-related areas of macaque cortex
uses the microcircuit model of \citep{Potjans14_785} as a prototype
for all 32 areas in the FV91 parcellation \citep{Felleman91_1} and
customizes it based on experimental findings on cortical structure.
From anatomical studies, it is known that cortical areas in the macaque
monkey are heterogeneous in their laminar structure and can be roughly
categorized into 8 different architectural types based on cell densities
and laminar thicknesses. This distinction was originally developed
for prefrontal areas \citep{Barbas97}, and then extended to the entire
cortex \citep{Hilgetag16_in_press}. The visual cortex, and thus the
model, comprises areas of categories 2, 4, 5, 6, 7 and 8. Precise
layer-specific neuron densities are available for a number of areas,
while for other areas, the neuron density is estimated based on their
architectural type (see \citep{Schmidt16_arxiv_v4} for details). 

The inter-areal connectivity is based on binary data from the CoCoMac
database \citep{Stephan01_1159,Bakker12_30,Felleman91_1,Rockland79_3,Barnes92_222}
indicating the existence of connections, and quantitative data from
\citep{Markov2014_17}. The latter are retrograde tracing data where
connection strengths are quantified by the fraction of labeled neurons
($FLN$) in each source area. The original analysis of the experimental
data was performed in the M132 atlas \citep{Markov2014_17}. Both
the FV91 and the M132 parcellations have been registered to F99 space
\citep{VanEssen02_574}, a standard macaque cortical surface included
with the software tool Caret \citep{VanEssen01_443}. This enables
mapping between the two parcellations. 

On the target side, we use the exact coordinates of the injections
to identify the equivalent area in the FV91 parcellation. To map the
data on the source side from the M132 atlas to the FV91 parcellation,
we count the number of overlapping triangles on the F99 surface
between any given pair of regions and distribute data proportionally
to the amount of overlap using the F99 region overlap tool on http://cocomac.g-node.org.
In the model, this $FLN$ is mapped to the indegree $K_{AB}$ the
target area $A$ receives from source area $B$ divided by its total
indegree, i.e., $FLN_{AB}=K_{AB}/\sum_{B^{\prime}}K_{AB^{\prime}}$.
Here, $K_{AB}$ is defined as the total number of synapses between
$A$ and $B$ divided by the total number of neurons in $A$. On the
source side, laminar connection patterns are based on CoCoMac \citep{Felleman91_1,Barnes92_222,Suzuki94_497,Morel90_555,Perkel86,Seltzer94}
and on fractions of labeled neurons in the supragranular layers ($SLN$)
\citep{Markov14}. Gaps in these data are bridged exploiting a sigmoidal
relation between $SLN$ and the logarithmized ratio of overall cell
densities of the two areas, similar to \citep{Beul15_arxiv}. We
map the $SLN$ to the ratio between the number of synapses originating
in layer 2/3 and the total number of synapses between the two areas,
assuming that only excitatory populations send inter-area connections,
i.e., $SLN_{AB}=\sum_{i}K_{AB}^{i,\mathrm{2/3E}}N_{A}^{i}/\sum_{i,j}K_{AB}^{ij}N_{A}^{i}$,
where the indices $i$ and $j$ go over the different populations
within area $A$ and $B$, respectively. In the context of the model,
we use the terms $FLN$ and $SLN$ to refer to the relevant relative
indegrees given here. On the target side, the CoCoMac database provides
data from anterograde tracing studies \citep{Felleman91_1,Rockland79_3,Jones78_291,Seltzer91_625}.

Missing inputs in the model, i.e., from subcortical and non-visual
cortical areas, are replaced by Poissonian spike trains, whose rate
$\nuext$ is a free, global parameter. In the original model all populations
of a particular area receive the same indegree of external inputs
$K_{\mathrm{ext}}$. The only exception to this rule is area TH where
the absence of granular layer 4 is compensated by an increase of the
external input to populations 2/3E and 5E by $20\,\%$. To elevate
the firing rates in the excitatory populations in layers 5 and 6,
we increase the external drive onto these populations. The possibility
of a higher drive onto these populations is left open by the sparseness
of the corresponding experimental data. We enhance the external Poisson
drive of the 5E population, parametrized by the $K_{5E,\mathrm{ext}}$
incoming connections per target neuron (indegree), in all areas by
increasing $\kappa=K_{\mathrm{5E,\mathrm{ext}}}/K_{\mathrm{ext}}$.
The simultaneous increase in the drive of 6E needs to be stronger,
since the firing rates in population 6E of the original model (\prettyref{fig:bistability-multiarea}C)
are even lower than the rates of 5E (averaged across areas: $0.09\spikess$
for 5E compared to $2\cdot10^{-5}\spikess$ for 6E). We thus scale
up $K_{\mathrm{6E,\mathrm{ext}}}$ linearly with $\kappa$ such that
$\kappa=1.15$ results in $K_{\mathrm{6E,\mathrm{ext}}}/K_{\mathrm{ext}}=1.5$.

\ifthenelse{\boolean{isarxiv}}{\section{Acknowledgement} The authors gratefully acknowledge the computing time granted (jinb33) by the JARA-HPC Vergabegremium and provided on the JARA-HPC Partition part of the supercomputer JUQUEEN  at Forschungszentrum J\"ulich. Partly supported by Helmholtz Portfolio Supercomputing and Modeling for the Human Brain (SMHB), the Helmholtz young investigator group VH-NG-1028, EU Grant 269921 (BrainScaleS). This project received funding from the European Union's Horizon 2020 research and innovation programme under grant agreement No. 720270. All network simulations carried out with NEST (http://www.nest-simulator.org).}{} 

\newpage{}\bibliographystyle{jneurosci_ab}

\end{document}